\documentclass[9pt, a4paper, twocolumn]{revtex4}
\usepackage{multirow}
\usepackage{amsmath}
\usepackage{graphicx}

\begin{document}

%\begin{frontmatter}

\title{Deconvolution of the High Energy Particle Physics Data with Machine Learning}


\author{Bora I\c{s}{\i}ldak}
\affiliation{Ozyegin University, Department of Natural and Mathematical Sciences, Istanbul, Turkey}
\affiliation{Middle East Technical University, Physics Department, Ankara, Turkey}
\author{Alper Hayreter}
\affiliation{Ozyegin University, Department of Natural and Mathematical Sciences, Istanbul, Turkey}
\affiliation{Middle East Technical University, Physics Department, Ankara, Turkey}
\author{Aidan R. Wiederhold}
\affiliation{University of Warwick,  Department of Physics, Coventry, U.K.}

\begin{abstract}
	A method for correcting smearing effects using machine learning technique is presented. Compared to the standard deconvolution approaches in high energy particle physics, the method can use more than one reconstructed variable to predict the value of unsmeared quantity on an event-by-event basis. In this particular study, deconvolution is interpreted as a classification problem, and neural networks (NN) are trained to  deconvolute the Z boson invariant mass spectrum generated with \textsc{MadGraph} and \textsc{pythia8} Monte Carlo event generators in order to prove the principle. Results obtained from the machine learning method is presented and compared with the results obtained with traditional methods.
\end{abstract}

\maketitle
%\begin{keyword}
%	Machine learning, Deep Learning, Unfolding
%\end{keyword}
%\end{frontmatter}

\section{Introduction}
\begin{figure}[!htbp]
	\begin{center}
		\includegraphics[width=\columnwidth]{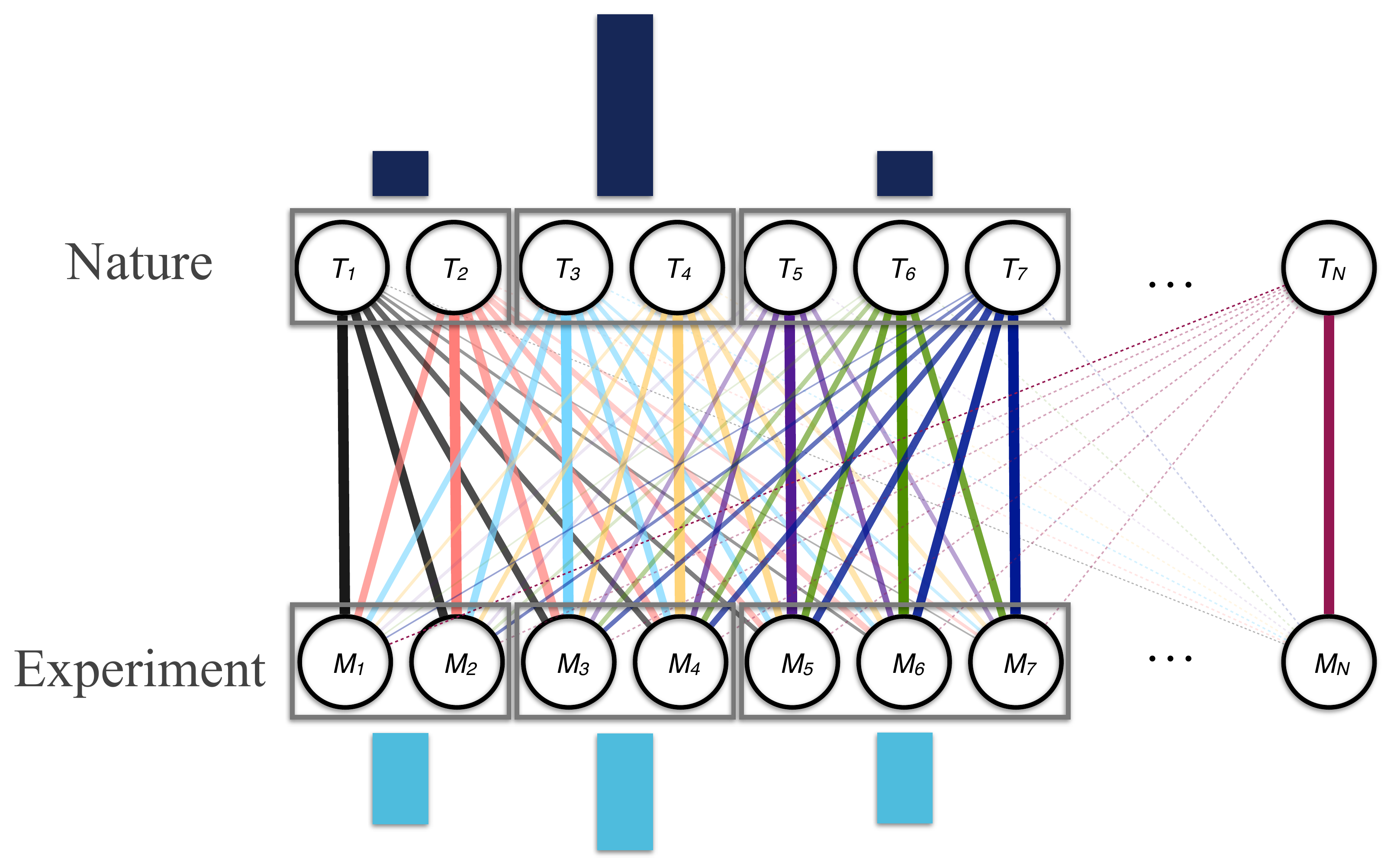}
		\caption{A simple illustration of the problem. Colored links from nature values to experiment values represent the related probabilities emerging from the resolution of the experimental apparatus.}
		\label{illustration}
	\end{center}
\end{figure}

In high energy particle physics (HEP), any measured spectrum (detector level) is a convoluted form of the true distribution (particle level) due to the several effects like trigger and event selection efficiencies, acceptance and finite resolution of the detectors. Actual value of the observable is measured as a different value with a probability associated to these smearing effects (see Figure \ref{illustration}). Measured spectrum can be related to the true spectrum by the following integral transformation,

\begin{equation} 
	N\left(y^{\textrm{m}}\right)=
	\int_{0}^{\infty} n\left(y^{\textrm{t}}\right) R\left(y^{\textrm{m}},~y^{\textrm{t}}\right) d y^{\textrm{t}}.
\end{equation}
Where $N(y^{\textrm{m}})$ is the number of events with $y^{\textrm{m}}$, $n(y^{\textrm{t}})$ is the number of events occurring with $y^{\textrm{t}}$ and $R(y^{\textrm{m}},~y^{\textrm{t}})$ is the probability of measuring an event as $y^{\textrm{m}}$ which is actually occurring with $y^{\textrm{t}}$.  Finding a proper deconvolution  (also called unfolding) for the resolution and efficiency effects to extract the underlying distribution is necessary, i.e. if one has to make a comparison between the spectra of same observable measured with different detectors or combine the spectra, again measured with different detectors. A common formulation of the problem is describing a linear transformation between the true and measured spectra since they can be represented discretely as vectors. This definition of the problem requires the inversion of the matrix that transforms true spectrum to measured spectrum.

\begin{equation}
	\mathbf{A}\vec{y}_{true}=\vec{y}_{measured},
\end{equation}

\begin{equation}
	\vec{y}_{true}=\mathbf{A}^{-1}\vec{y}_{measured}.
\end{equation}

Simple inversion methods may lead to significant negative correlations among the bins of the deconvoluted distribution, hence may not give robust results. Several  matrix regularization methods have been developed and used to overcome this problem. Also, iterative methods applying successive convolutions to obtain zeroth convolution by extrapolation have been suggested. An extensive comparison of the existing methods can be found in Ref. \cite{Schmitt-Review}. 

In this letter, deconvolution is defined as a multi-class categorization problem where the categories represent the correct bin that a measured value should actually be placed. A recent work shows that such ML approach to deconvolution problem can give promising results even with the predictions relying on single input Ref. \cite{Glazov}. Modern machine learning (ML) methods are highly advanced in categorization problems. There are two main types of categorization problems: multi-class categorization and multi-label categorization. In multi-class categorization, each data point can be assigned with one class where there can exist at least two different classes. For example, particle collision data can be categorized as a ``background event" or a ``signal event" depending on the kind of interested physics process defined as the signal. On the other hand, multi-label categorization allows multiple assignments of the labels. A number of ML methods has been developed to solve categorization problems such as linear discriminators, nearest neighborhood relations,  decision trees, neural networks etc. Among these methods, neural networks has a vast potential for solving even the most complex problems in exchange of computational resources. Developments in the CPU technology enhanced the interest in neural networks, and a significant progress has been achieved in the recent decade.  

\section{Methodology}
\subsection{Sample Simulation and Z Mass Reconstruction}
\begin{figure}[!htbp]
	\begin{center}
		\includegraphics[width=\columnwidth]{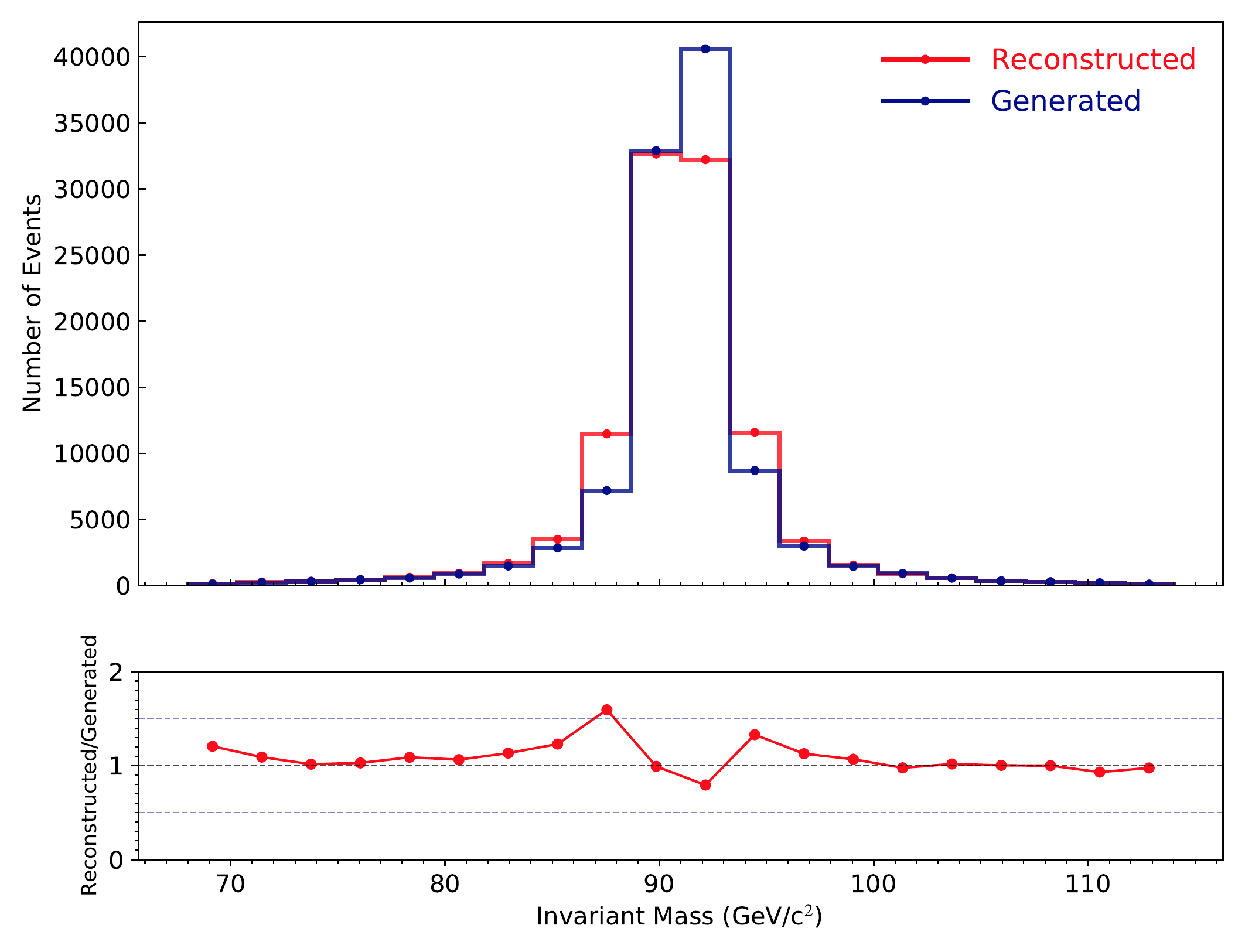}
		\caption{Histograms showing the generated and reconstructed $Z$ boson mass values in 20 bins. The effect of the detector resolution can easily be seen by looking at the difference between two spectra.}
		\label{m_reco_gen}
	\end{center}
\end{figure}

In order to test the proposed ML deconvolution techniques with a realistic scenario, a sample containing 500k  $pp\rightarrow Z$ events at 13 TeV is produced by using \textsc{MadGraph}5 \cite{MG5} event generator where $Z$ boson subsequently decaying into opposite charge lepton pairs ($\ell^+\ell^-$ or $\mu^+\mu^-$). The \textsc{pythia8} \cite{pythia8} event generator is employed to perform parton showering and hadronization. As a final stage, Delphes \cite{Delphes} fast detector simulation is used to get detector-like physics outputs, i.e. tracks, electrons, jets, missing transverse energy (MET). In order to reconstruct the $Z$ boson, invariant masses of all possible pairs of oppositely charged, same flavor isolated leptons are calculated. The  pair that gives the least absolute difference with the $Z$ boson mass value taken from the Particle Data Group (PDG)  \cite{PDG} is accepted  as the $Z$ candidate mass ($m_Z^{\textrm{rec}}$) if it lies within $\pm$25\% vicinity of the $m_Z^{\textrm{pdg}}$. Each lepton is required to be isolated where the isolation criterion is the scalar sum of the transverse momenta of the particle tracks (apart from the lepton itself) in a cone around its direction should be less than 10\% of the lepton $p^T$. For $\sim$225k events satisfying the reconstruction criteria, the true value of the $Z$ boson mass ($m_Z^{\textrm{gen}}$) is calculated by using the generator level information stored in the Delphes output. Two histograms with 20, 40 and 60 uniform bins between 65 GeV and 115 GeV are filled with the values of $m_Z^{\textrm{gen}}$ and $m_Z^{\textrm{reco}}$, see Figure \ref{m_reco_gen}. It can be clearly seen that the resolution effects are highly pronounced on the reconstructed spectrum. 
\subsection{Input Selection and Model Training}
The $p^T$, $\eta$, $\phi$, isolation values and PDG identification numbers (PDGID) of the leptons in the $Z$ decay candidate pair are stored with the invariant mass of the pair as well, later to be used as the input features of the ML training. Input features are ranked by using scikit-learn \cite{scikit-learn} feature selection class. Naturally, the most important feature is the reconstructed mass value $m_Z^{\textrm{rec}}$. According to the feature ranking, the five most important input features are kept for the training, namely, $m_Z^{\textrm{reco}}$, $p^T$ and isolation variable of the daughter leptons. After standardized by removing the mean and scaling to unit variance, the data are split into training and test subsets with 9:1 ratio. Two different sequentially layered neural networks are trained by using the input data described above, and the performance is compared with the result obtained with TUnfold tool \cite{TUnfold}. Table \ref{table:bin_classification_summary} shows the structure of the trained network.

In this multi-class categorization approach, the bin numbers of the generated mass values are given as the target categories where there are as many target categories as the number of bins. The training converges to give a model that predicts the correct bin. Hence, the model gives a multi-dimensional vector with associated probabilities for each event. Instead of filling the histogram by taking the predicted bin number with the highest probability, all possible bins are filled with weights according to the predicted probabilities. More explicitly, the number of events in the $j^{\textrm{th}}$ bin has been calculated in terms of the prediction probabilities as follows;

\begin{equation} \label{discreteConvolution}
	N_j = \sum_{k=1}^{n_{\textrm{events}}}\sum_{i=1}^{n_{\textrm{bins}}} p_k^i \delta_{ij}
\end{equation}

\noindent where $p_k^i$ is the predicted probability of falling within the $i^\textrm{th}$ bin for the $k^{\textrm{th}}$ event. Basically,  Equation. \ref{discreteConvolution} is a discrete event–by–event basis convolution of the reconstructed data with the ML predictions such that it gives a close approximation to the generated distribution. A reduced version of the problem by calculating the migration distance of each reconstructed mass value in units of bin widths is also studied. Such reduction showed that ~95\% of the bin migrations are within $\pm$1 bin. However, this kind of reduction of the migration range depends on the number of bins to be used. The migration range should be extended with the increasing number of bins. Similarly, the sum of probabilities is used for histogram reconstruction, however the bins filled are determined by adding each migration vector in turn to the bin number occupied by the reconstructed mass value for the event. 

\begin{table}[!htbp]
	\resizebox{0.99\columnwidth}{!}{
		\centering
		\def\arraystretch{1.1}
		{\begin{tabular}{|c|c|c|}
				\hline
				Layer Number  	&  Layer Type & Layer Specific Details 	\\\hline\hline
				0  & Input & Input Dimension: Number of kept features, Output Dimension: 100, \\
				& &  Activation: Linear \\\hline
				1  & Dense & Output Dimension: $\alpha N^p$: $N=$ number of bins,, Activation: ReLU\\\hline
				2 & Output & Output Dimension: Number of bins, Activation: Softmax \\\hline
	\end{tabular}}}
	\caption{Summary of the model used to perform bin classification. For 20 bins training , $\alpha=8$ and $p=2$ are used as in Ref. \cite{Glazov}. For 40 and 60 bins trainings , $\alpha=1$ and $p=1$ are used to decrease training time.}
	\label{table:bin_classification_summary}
\end{table}
\begin{table}[!hbpt]
	\resizebox{0.99\columnwidth}{!}{
		\centering
		{\begin{tabular}{|c|c|c|}
				\hline
				Layer Number  	&  Layer Type & Layer Specific Details 	\\\hline\hline
				0  & Input & Input Dimension: Number of kept features, Output Dimension: 100, \\
				&       & Kernel Initializer: Glorot normal, Activation: Linear \\\hline
				1  & Dense & Output Dimension: 600, Activation: ReLU \\\hline
				2  & Batch Normalization &  \\ \hline
				3  & Dropout & Ratio: 0.1, Seed: 23 \\\hline
				4  & Dense   & Output Dimension: 750, Activation: Tanh \\\hline
				5  & Batch Normalization &  \\ \hline
				6  & Dropout & Ratio: 0.1, Seed: 46 \\\hline
				7  & Dense   & Output Dimension: 600, Activation: ReLU \\\hline
				8  & Batch Normalization &  \\ \hline
				9  & Output & Output Dimension: Number of classes, Activation: Softmax \\\hline
	\end{tabular}}}
	\caption{Summary of the model used to perform migration vector classification.}
	\label{table:migvecClassificationSummary}
\end{table}

\section{Results and Discussion}
Although ML training takes much more time compared to the TUnfold or any other standard deconvolution method, modern ML tools makes this time reasonable even for personal computers. For example, the training of one epoch of the categorization training  takes $\sim$5 seconds with 2.2 GHz Intel Core i7 processor using with Keras using the Tensorflow \cite{tensorflow} backend with GPU acceleration.   

\begin{figure}[h!]
	\begin{center}
		\includegraphics[width=\columnwidth]{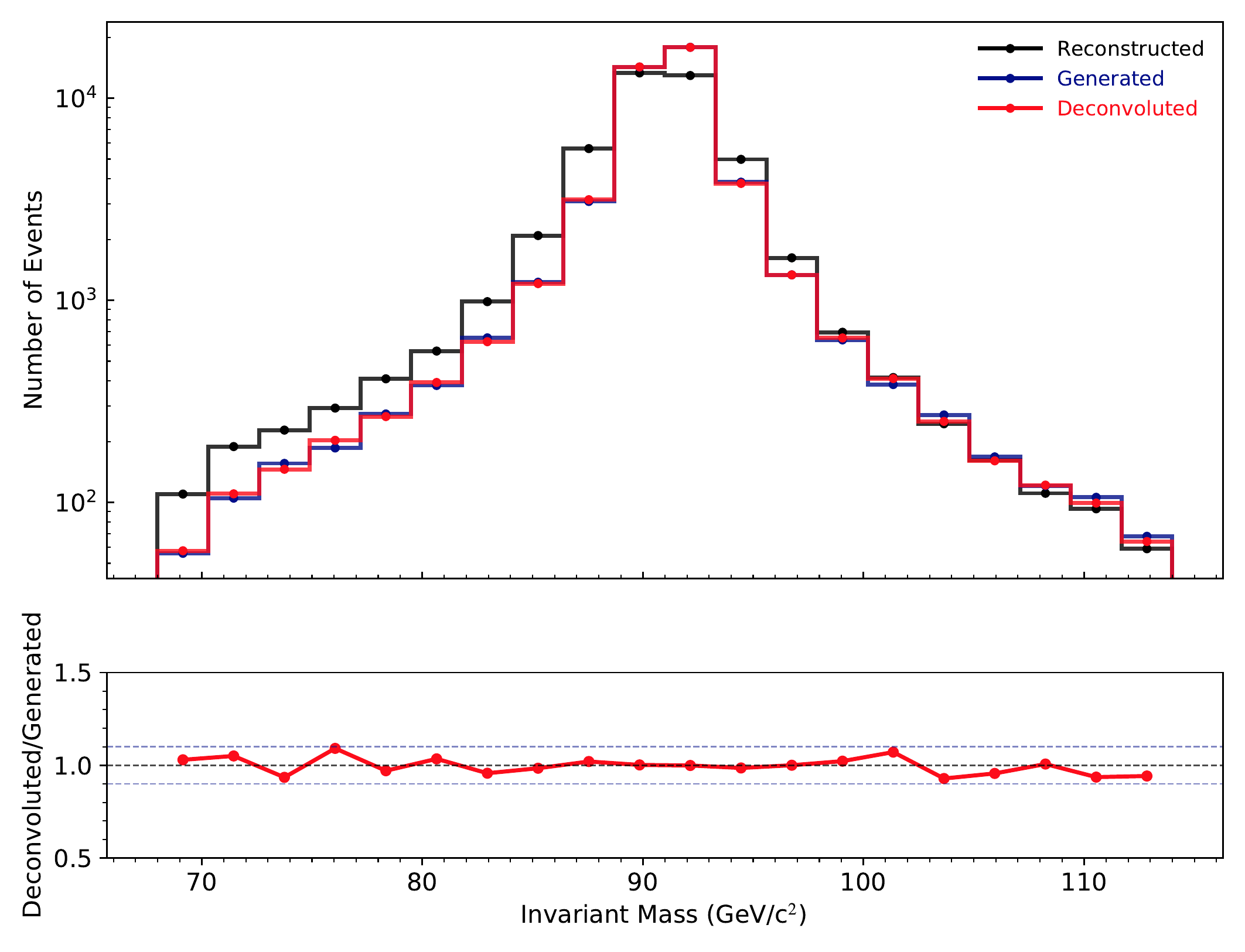}
		\includegraphics[width=\columnwidth]{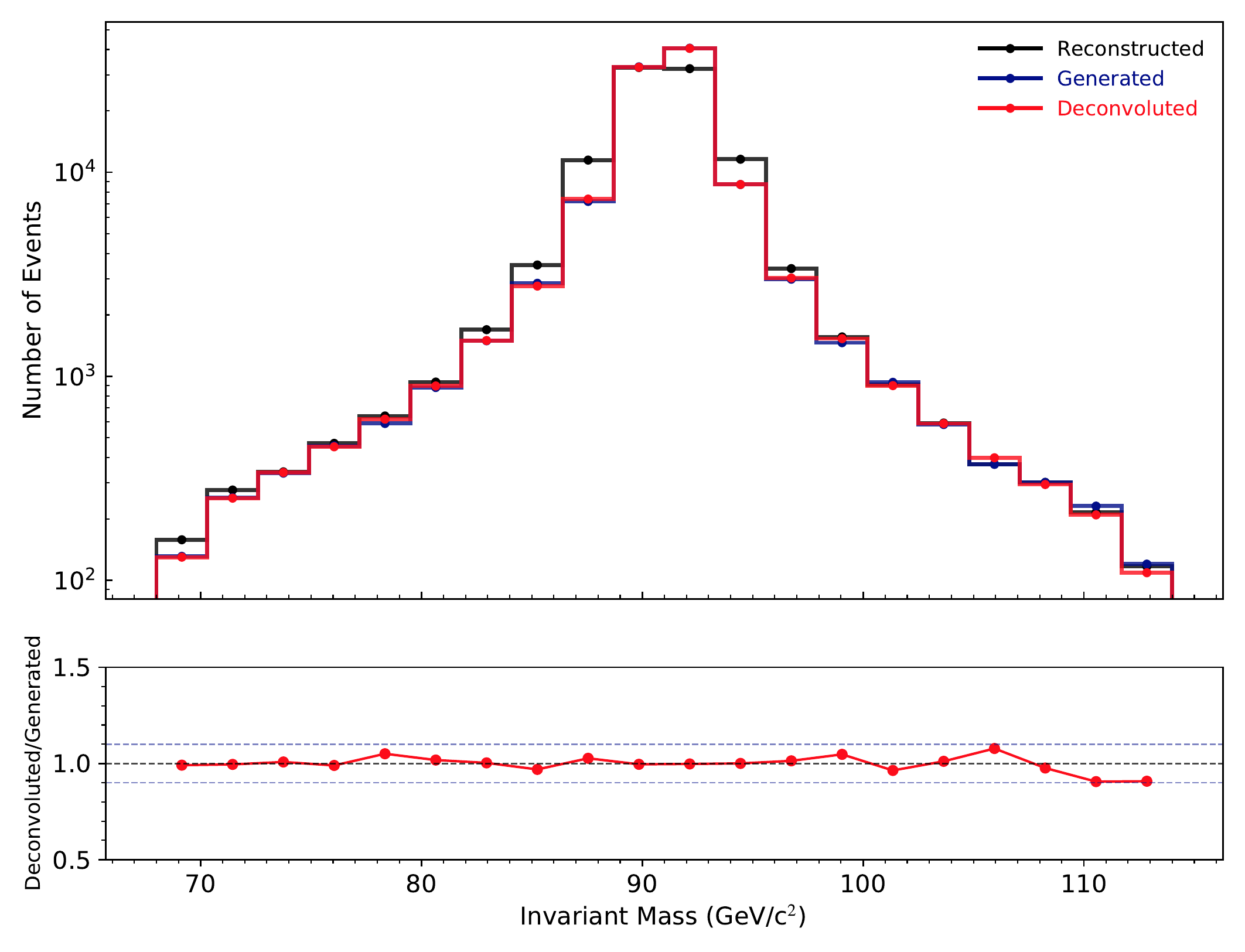}
	\end{center}
	\caption{ML deconvolution results on the test sample for both bin (top plot) and migration (bottom plot) classification interpretations.}
	\label{fig:categorization_results}
\end{figure}

\begin{figure}[!hbpt]
	\begin{center}
		\includegraphics[width=\columnwidth]{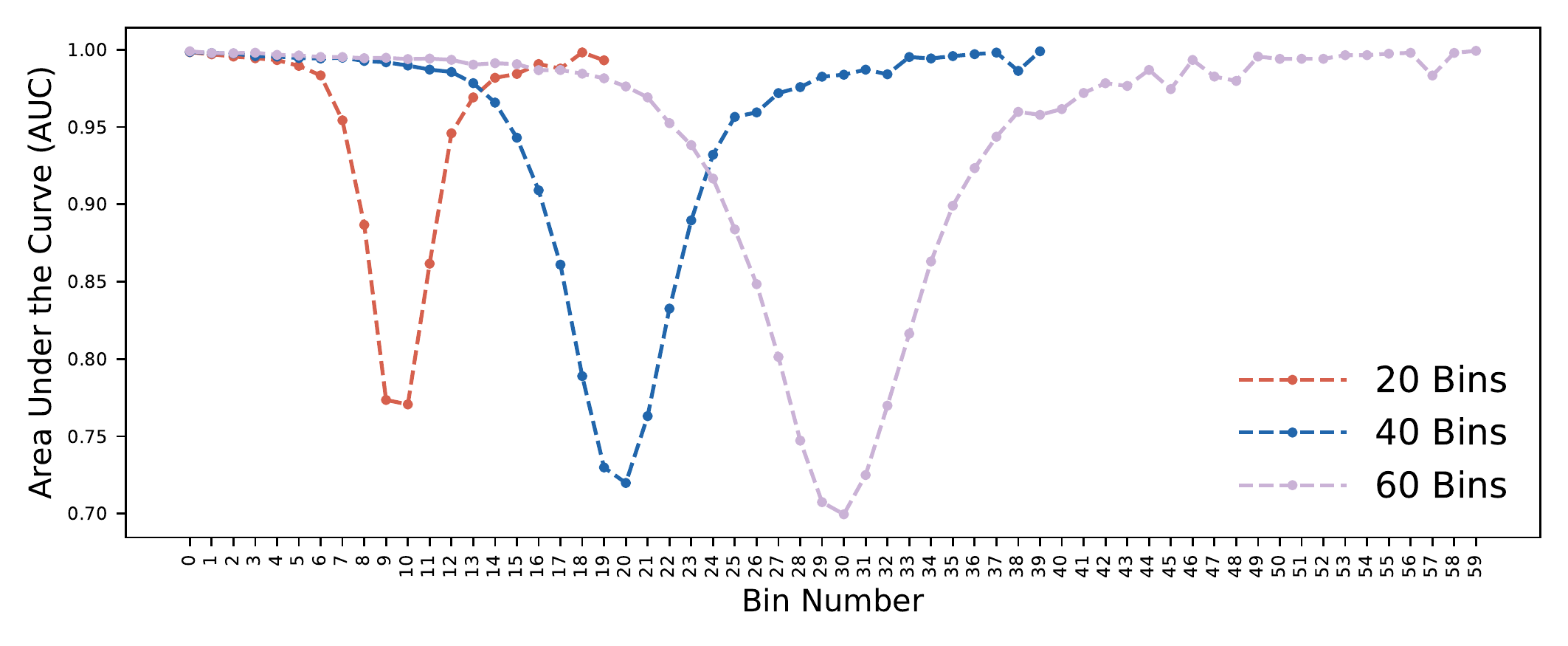}
	\end{center}
	\caption{AUC scores for all 20, 40 and 60 classes (bins) with one-versus-all approach.}
	\label{fig:auc}
\end{figure}

Model training is performed on $\sim$180k of simulated events with a batch normalization \cite{batch_normalization} over every 1024 events for 200 epochs. Categorical cross-entropy \cite{Goodfellow} with the ‘Adam’ optimizer \cite{Adam} which employs an adaptive learning rate is used as the loss function. Mathematically, categorical cross-entropy is given as the following equation:

\begin{equation} \label{CCEloss}
	-\sum_{c=1}^{N} y_{i,c} \log(p_{i,c})
\end{equation}

\noindent where $N$ is the total number of classes, $y_{i,c}$ is a binary indicator (0 or 1) that indicates whether $c$ is the correct class. For reducing the risk of overfitting, an early stopping callback is implemented that terminates the training if the improvement stays within a range for a given number of epochs. Figure \ref{fig:categorization_results} shows the resulting deconvoluted histograms for two different interpretation of the bin classification problem.
%The agreement between the generated and deconvoluted spectra seems quite well. 
Although there is no perfect performance metric for the multi-class categorization models, area under the curve (AUC) values give the separation power of the model for each class. Figure \ref{fig:auc} shows the AUC scores calculated with one-versus-all approach for 20, 40 and 60 bins. In all these different binning cases they cover 65-115 GeV/c$^2$ range.

\begin{figure}[ht]
	\begin{center}
		\includegraphics[width=\columnwidth]{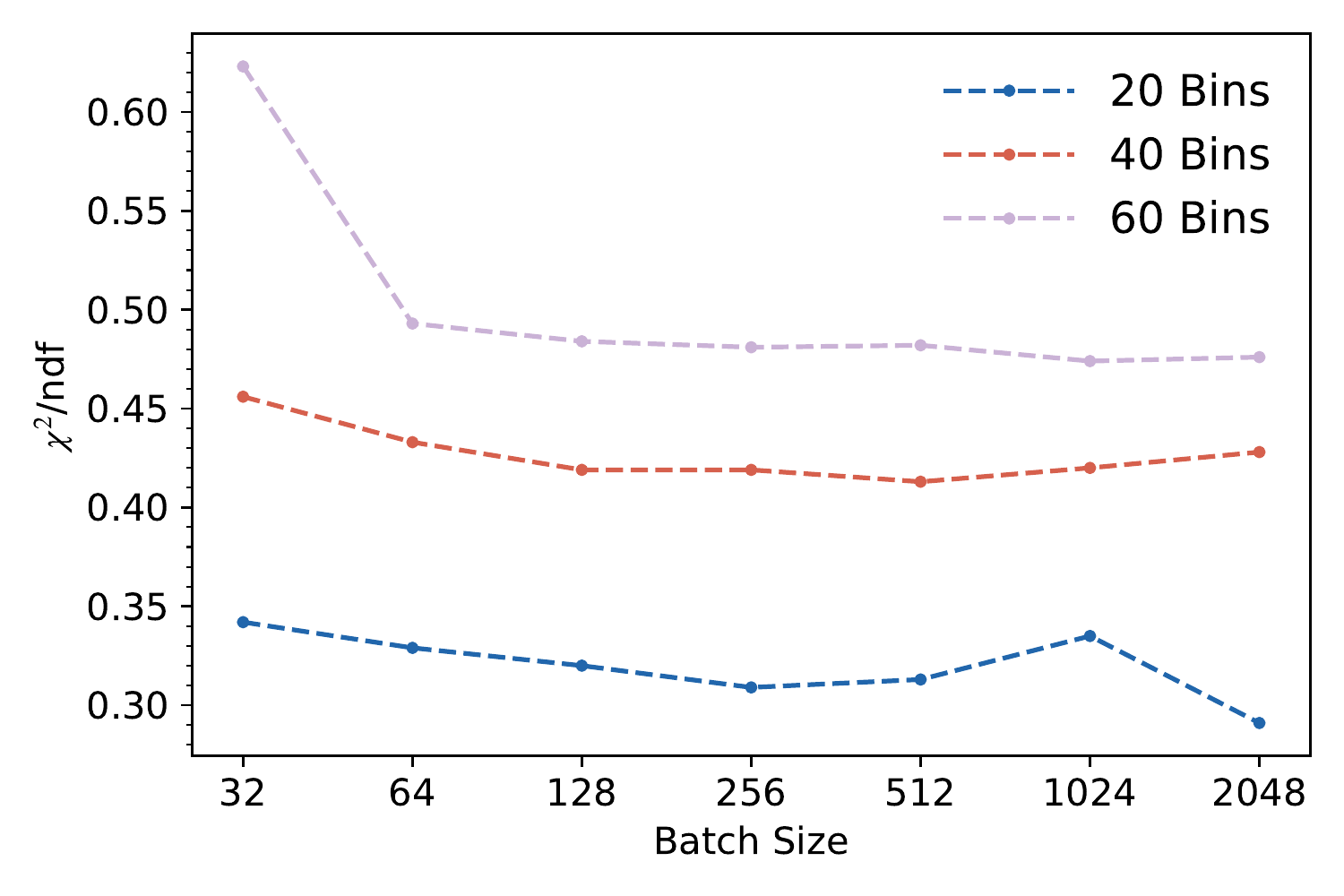}
		\includegraphics[width=\columnwidth]{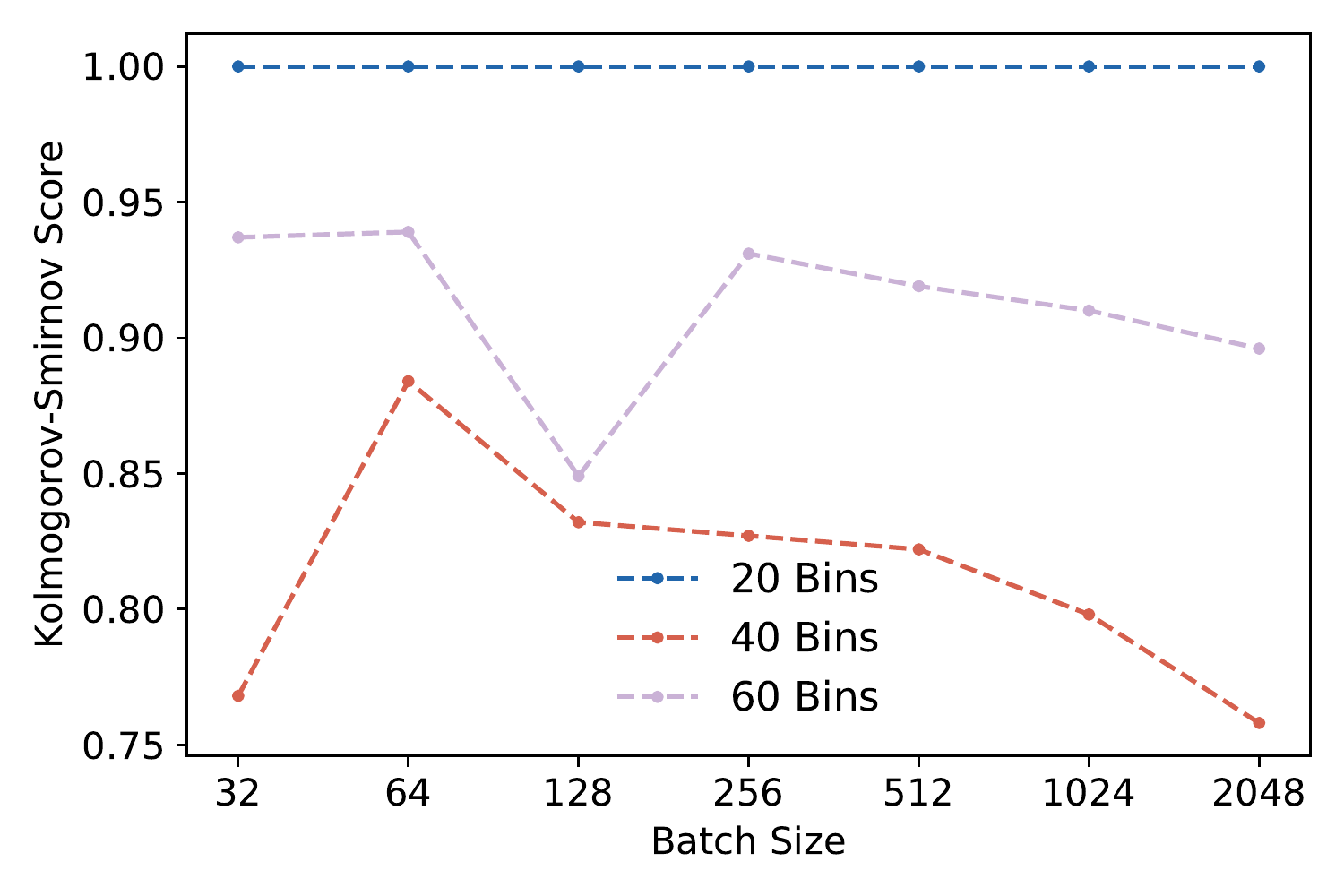}
		\includegraphics[width=\columnwidth]{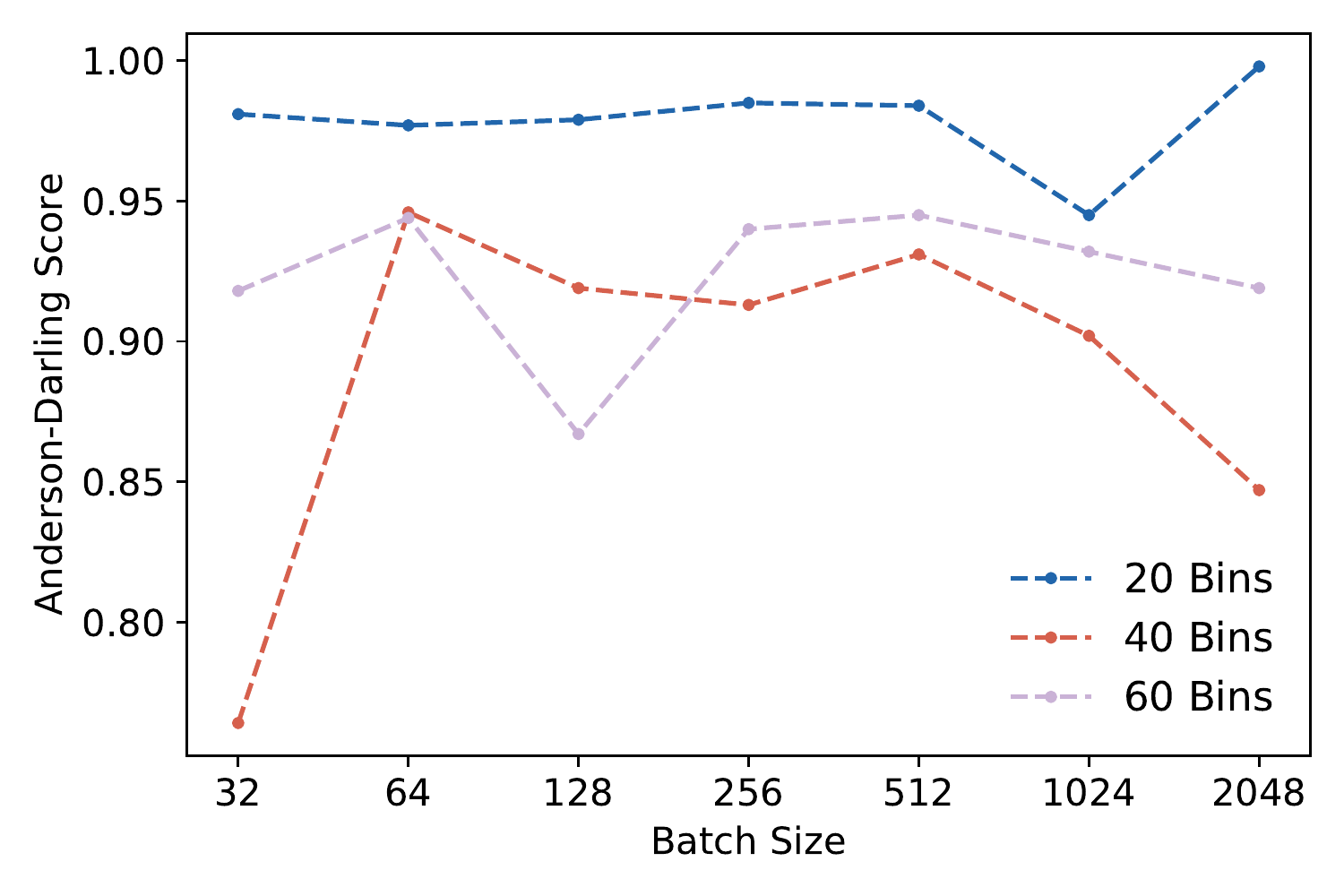}
	\end{center}
	\caption{ML migration categorization result for the test split.}
	\label{fig:batchsize_dependence}
\end{figure}

Nevertheless, the most important measure of the model performance for this study is the agreement between generated and deconvoluted spectra. It can be seen from Figure \ref{fig:categorization_results} that the deconvoluted histogram is in very good agreement with the generated one. The ratio plot on the bottom panel of the figure shows that their bin entries differ below 10\%. Furthermore, $\chi^2$/ndof, Anderson-Darling \cite{AD} and Kolmogorov-Smirnov tests are applied to quantify the agreement. Also, to check the robustness of the methods, a 10-fold cross-validation is performed for 20, 40, 60, 80 and 100 bins. The results of the corresponding  $\chi^2$/ndof, Anderson-Darling and Kolmogorov-Smirnov tests are summarized in Table \ref{table:testscores}. It is found that the accuracy is not a good predictor of whether the final deconvoluted histogram agrees with the generated histogram or not. For example, models for direct bin classification with 20 bins reaches up to $\sim 55 \%$ accuracy both in training and test samples whereas the resulting deconvoluted histogram ends up with a Kolmogorov--Smirnov Test score $\sim 1$, an Anderson--Darling Test score $\sim 1$ and a $\chi^2$/ndof value less than unity. Another hyper-parameter that might effect the performance is the batch size used in the training. It is known that using smaller batch sizes has a regularizing effect and it offers a better generalization. The same network is trained for different batch sizes for 20, 40 and 60 bins. Figure \ref{fig:batchsize_dependence} shows that using different batch sizes has no considerable effect on the deconvolution performance. Although the test accuracy is almost identical to the model's accuracy trained with the categorical cross entropy loss function, it gives unacceptable $\chi^2$/ndof, Anderson-Darling and Kolmogorov-Smirnov test scores.

\begin{figure}[ht]
	\begin{center}
		\includegraphics[width=\columnwidth]{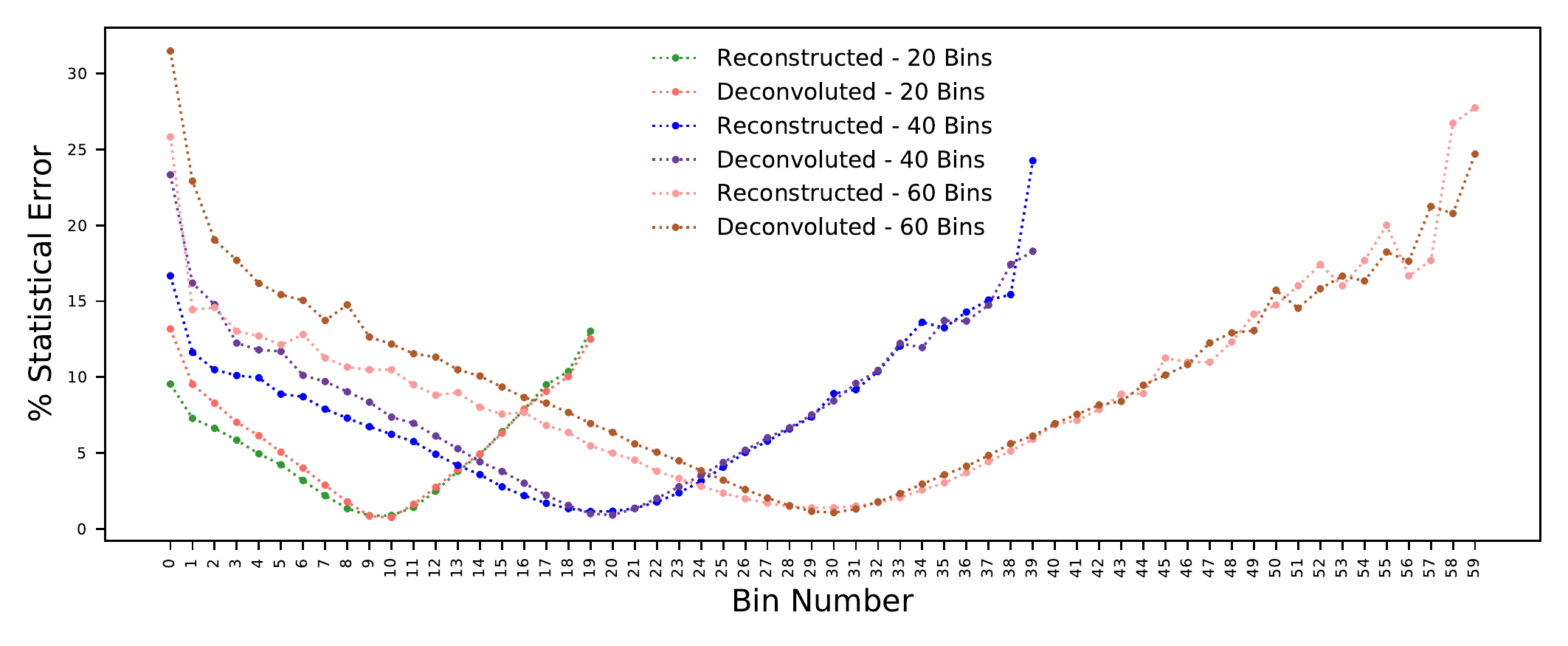}
	\end{center}
	\caption{ Fractional statistical uncertainties of reconstructed and deconvoluted spectra for 20, 40 and 60 bins.}
	\label{fig:stat_errors}
\end{figure}

\begin{table}[!hbpt]
\resizebox{0.99\columnwidth}{!}{
		\centering
		{\begin{tabular}{|c|c|c|c|c|c|}
				\hline
				\multirow{2}{*}{Method} & \multicolumn{5}{|c|}{Kolmogorov-Smirnov Test Scores} \\ \cline{2-6}
				& 20 Bins & 40 Bins & 60 Bins & 80 Bins & 100 Bins \\
				\hline \hline
				ML & 1.000 $\substack{+0.000 \\ -0.000}$ & 0.925 $\substack{+0.075 \\ -0.223}$ & 1.000 $\substack{+0.000 \\ -0.000}$ & 0.956 $\substack{+0.044 \\ -0.131}$ & 1.000 $\substack{+0.000 \\ -0.000}$ \\
				\hline
				TUnfold & 0.898$ \substack{+0.102 \\ -0.193}$ & 1.000 $\substack{+0.000 \\ -0.000}$ & 0.998 $\substack{+0.002 \\ -0.006}$ & 0.986 $\substack{+0.014 \\ -0.029}$ & 0.997 $\substack{+0.003 \\ -0.010}$\\
				\hline
	\end{tabular}}}
	\\
	\resizebox{0.99\columnwidth}{!}{
		\centering
		\begin{tabular}{|c|c|c|c|c|c|}
			\hline
			\multirow{2}{*}{Method} & \multicolumn{5}{|c|}{$\chi^2$/ndf Values} \\ \cline{2-6}
			& 20 Bins & 40 Bins & 60 Bins & 80 Bins & 100 Bins \\
			\hline
			\hline
			ML & 0.064 $\substack{+0.033 \\ -0.033}$& 0.120 $\substack{+0.181 \\ -0.120}$ & 0.018 $\substack{+0.007 \\ -0.007}$ & 0.091 $\substack{+0.130 \\ -0.091}$ & 0.036 $\substack{+0.028 \\ -0.028}$ \\
			\hline
			TUnfold & 1.390 $\substack{+0.550 \\ -0.055}$ & 1.010 $\substack{+0.340 \\ -0.340}$ & 1.030 $\substack{+0.300 \\ -0.300}$ & 1.020 $\substack{+0.180 \\ -0.180}$ & 1.110 $\substack{+0320 \\ -0.320}$ \\
			\hline
	\end{tabular}}
	\resizebox{0.99\columnwidth}{!}{
		\centering
		\begin{tabular}{|c|c|c|c|c|c|}
			\hline
			\multirow{2}{*}{Method} & \multicolumn{5}{|c|}{Anderson Darling Test Scores} \\ \cline{2-6}
			& 20 Bins & 40 Bins & 60 Bins & 80 Bins & 100 Bins \\
			\hline
			\hline
			ML & 0.988 $\substack{+0.012 \\ -0.030}$ & 0.905 $\substack{+0.095 \\ -0.272}$ & 1.00 $\pm$ 0.00 & 0.935 $\substack{+0.065 \\ -0.189}$ & 0.999 $\substack{+0.001 \\ -0.004}$ \\
			\hline
			TUnfold & 0.00 $\pm$ 0.00 & 0.00 $\pm$ 0.00 & 0.00 $\pm$ 0.00 & 0.00 $\pm$ 0.00 & 0.00 $\pm$ 0.00 \\
			\hline
	\end{tabular}}
	\caption{Kolmogorov-Smirnov, $\chi^2$/ndof and Anderson-Darling \cite{AD} scores from 10-fold cross-validation.}
	\label{table:testscores}
\end{table}

In the case of migration classification, the migration range on the training data is set by removing events migrated from outside the specified range. Decreasing the total number of classes the model has to learn could result in a higher overall accuracy since the model can learn about each class better. However, the model performance must be tested on a  sample which is uncut. The test sample which has had events removed for this reason will be referred as the ``cut data'' and the data in which some of these events have been added back will be referred as ``uncut data''. Table \ref{table:percentage_effect} illustrates the effect of setting a migration range on the performance of the method. The second column states the percentage of events that migrated by at most $n$ bins, where $n$ is given in the first column. Despite the model performs well on the cut sample (Table \ref{table:percentage_effect} bottom plot), it gives unacceptable evaluation scores for the uncut one. If the migration range is set in a way that no data remains outside, then the model ends up with successful scores. However, this often results in having more classes than the direct bin classification approach. Also, the network to be used is usually deeper than that is to be used for the bin classification, and takes longer training time. Furthermore, there is no guarantee that the result will be better or even as good as one obtained from bin classification. Therefore, this method, in its current state, offers no advantage over bin classification.

\begin{figure}
	\vspace*{0.5cm}
	\begin{center}
		\includegraphics[width=\columnwidth]{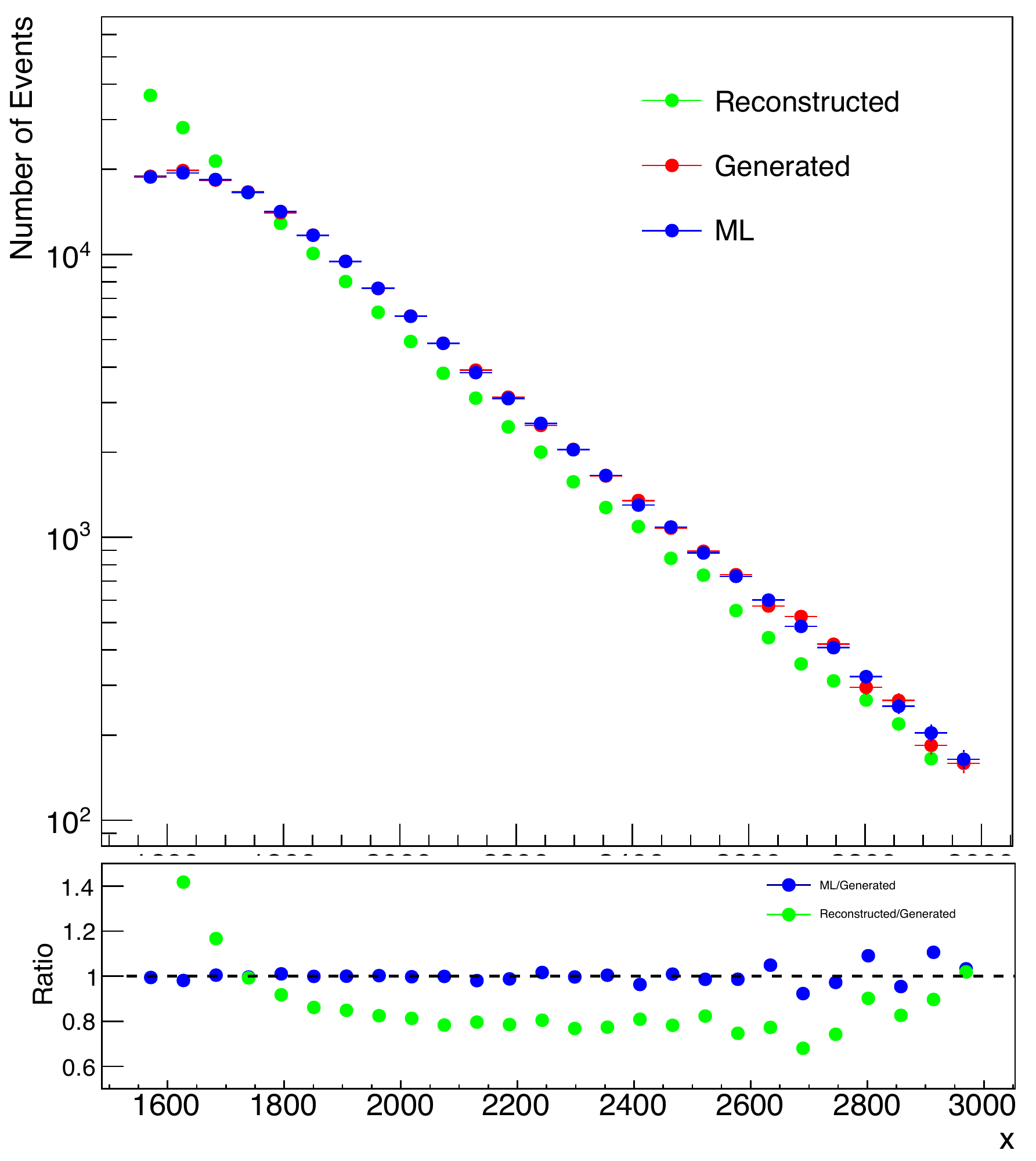}
	\end{center}
	\caption{ML results for a steeply falling toy distribution.}
	\label{fig:dijet}
\end{figure}

In order to see the effect of the underlying shape, another sample is produced by randomly generating 1M values, $x$,  in the range $[1500, 3000]$ following the shape $a(1-x)^b/x^c$, where $a, b, c > 0$ are inspired by \cite{dijetPaper}. To obtain the ``measured'' values, generated $x$ values are smeared using a gaussian distribution with relative standard deviations equal to $0.03x$ and $0.10x$ separately. Then the network described in Table \ref{table:bin_classification_summary} is again trained for deconvolution. The ``generated", ``reconstructed" and the ``deconvoluted (ML)" spectra are shown in Figure \ref{fig:dijet}. $\chi^2$/ndof, Anderson-Darling and Kolmogorov-Smirnov test scores are given in Table \ref{table:dijetScores}.

\begin{table}[!ht]
	\resizebox{0.99\columnwidth}{!}{
		\centering
		\begin{tabular}{|c|c|c|c|c|c|c|}
			\hline
			\multirow{2}{*}{Method} & \multicolumn{2}{|c|}{KS Test} & \multicolumn{2}{|c|}{$\chi^2$/DOF} & \multicolumn{2}{|c|}{AD Test} \\ \cline{2-7}
			&$\sigma=0.03$ &$\sigma=0.10$ &$\sigma=0.03$ &$\sigma=0.10$&  $\sigma=0.03$ &  $\sigma=0.10$ \\
			\hline
			\hline
			ML   & 0.981 & 1.00 & 0.491 & 0.824 & 0.991 & 0.985\\
			\hline
	\end{tabular}}
	\caption{Test scores obtained by applying bin classification to a steeply falling toy distribution (60 bins)}
	\label{table:dijetScores}
\end{table}

The advantage of this event-by-event deconvolution method is that the final spectrum is immune to the bin-to-bin correlations, a common problem of the matrix inversion based deconvolution techniques. Hence, the statistical uncertainties of the deconvoluted spectrum are calculated as the square root of the bin content. Figure \ref{fig:stat_errors} shows the fractional statistical uncertainties of both reconstructed and deconvoluted spectra in each bin for 20, 40 and 60 bins deconvolutions. It can be deduced from the figure that statistical uncertainties on the deconvoluted spectrum is systematically higher than the reconstructed one, especially on the left hand side of the spectrum. 

In conclusion, it is shown that bin to bin classification with an alternate interpretation of the predictions is a powerful method for histogram deconvolution that works for a range of total bin numbers and two different distribution shapes.

\begin{table}[!hbpt]
	\resizebox{0.99\columnwidth}{!}{
		\centering
		\begin{tabular}{|c|c|c|c|c|c|}
			\hline
			Migrations & \% of Data & Cut KS & Cut $\chi^2$/ndof & Uncut KS & Uncut $\chi^2$/ndof \\\hline\hline
			$\pm$ 1 & 91.5 & 0.902 & 0.775 & 0.00 & 20.0 \\\hline
			$\pm$ 2 & 97.8 & 0.673 & 1.03 & 0.00 & 8.30  \\\hline
			$\pm$ 3 & 98.8 & 1.00 & 0.558 & 0.00 & 5.49  \\\hline
			$\pm$ 4 & 99.1 & 0.391 & 1.99 & 0.00 & 6.23  \\\hline
			$\pm$ 5 & 99.4 & 0.998 & 0.686 & 0.0986 & 3.40 \\\hline
			$\substack{+17 \\ -5}$ & 100 & 0.914 & 1.25 & 0.914 & 1.25 \\\hline
	\end{tabular}}
	\caption{Performance comparison (on a 20 bin histogram) between reconstructing cut data and uncut data.}
	\label{table:percentage_effect}
\end{table}

%\input{AppendixA.tex}
%\section{Acknowledgments}
%This work is supported by CERN Summer Student Program and Turkish Atomic Energy Authority (TAEK) under the project 2018TAEK(CERN)A5.H6.F2-16. We would like to express our gratitude to  Mehmet Zeyrek for his invaluable support, to Stefan Schmitt and Bugra Bilin  for their insightful suggestions. 

%\bibliography{Untitled.bib}

\begin{thebibliography}{1}
	\bibitem{Schmitt-Review} Schmitt, S. Data Unfolding Methods in High Energy Physics. EPJ Web Conf. \textbf{137}, 11008 (2017).\\

	\bibitem{Glazov} Alexander, G Machine learning as an instrument for data unfolding. arXiv. 1712.01814 (2017).\\
	
	\bibitem{MG5} J. Alwall, R. Frederix, S. Frixione, V. Hirschi, F. Maltoni, O. Mattelaer, H. S. Shao, T. Stelzer, P. Torrielli, and M. Zaro, The automated computation of tree-level and next-to- leading order differential cross sections, and their matching to parton shower simulations, JHEP, 07, 2014.\\
	
	\bibitem{pythia8} T. Sj\"{o}strand, S. Mrenna and P. Skands, A brief introduction to PYTHIA 8.1. Comput. Phys. Comm. 178, 2008.\\
	
	\bibitem{Delphes} The DELPHES 3 collaboration, J. de Favereau, C. Delaere, P. Demin, A. Giammanco, V. Lema\^itre, A. Mertens, and M. Selvaggi, Delphes 3: a modular framework for fast simulation of a generic collider experiment. Journal of High Energy Physics, 2014.\\
	
	\bibitem{PDG} Particle Data Group (PDG), Review of particle physics, Phys. Rev. D, 98 (2018), p. 030001.\\
	
	\bibitem{scikit-learn} F. Pedregosa, G. Varoquaux, A. Gramfort, V. Michel, B. Thirion, O. Grisel, M. Blon- del, P. Prettenhofer, R. Weiss, V. Dubourg, J. Vanderplas, A. Passos, D. Cournapeau, M. Brucher, M. Perrot, and E. Duchesnay, Scikit-learn: Machine learning in Python, Journal of Machine Learning Research, 12 (2011), pp. 2825–2830.\\
	
	\bibitem{TUnfold} S. Schmitt, TUnfold, an algorithm for correcting migration effects in high energy physics, Journal of Instrumentation, 7 (2012), pp. T10003–T10003.\\
	
	\bibitem{tensorflow} M. Abadi et al., Tensorflow: Large-scale machine learning on heterogeneous distributed systems, CoRR, abs/1603.04467 (2016).\\
	
	\bibitem{batch_normalization} S. Ioffe and C. Szegedy. Batch normalization: Accelerating deep network training by reducing internal covariate shift. pages 448–456, 2015.\\
	
	\bibitem{Goodfellow} Goodfellow, Y. Bengio, and A. Courville, Deep Learning. The MIT Press, 2016.\\
	
	\bibitem{Adam} D. P. Kingma and J. Ba, Adam: A method for stochastic optimization, CoRR, abs/1412.6980 (2014).\\
	
	\bibitem{AD} M. A. Stephens, Edf statistics for goodness of fit and some comparisons, Journal of the American Statistical Association, 69 (1974), pp. 730–737.\\
	
	
	\bibitem{dijetPaper} CMS Collaboration, Search for narrow resonances and quantum black holes in inclusive and b-tagged dijet mass spectra from pp collisions at  $\sqrt{s} = 7$ TeV, JHEP 01 (2013) 013, doi:10.1007/JHEP01(2013)013, arXiv:1210.2387.
	
\end{thebibliography}

\end{document}